\documentclass[pra,twocolumn,amsfonts,amsmath,showpacs]{revtex4-1}
\usepackage{graphicx}
\begin{document}
	\title{Efficient algorithm for generating Pauli coordinates for an arbitrary linear operator}
	
	\author{Daniel Gunlycke}
	\email{daniel.gunlycke@nrl.navy.mil}
	\affiliation{U.S. Naval Research Laboratory, Washington, DC 20375, United States}
	\author{Mark C. Palenik}
	\affiliation{U.S. Naval Research Laboratory, Washington, DC 20375, United States}
	\author{Alex R. Emmert}
	\affiliation{SEAP Student, U.S. Naval Research Laboratory, Washington, DC 20375, United States}
	\author{Sean A. Fischer}
	\affiliation{U.S. Naval Research Laboratory, Washington, DC 20375, United States}
	
	\begin{abstract}
		Several linear algebra routines for quantum computing use a basis of tensor products of identity and Pauli operators to describe linear operators, and obtaining the coordinates for any given linear operator from its matrix representation requires a basis transformation, which for an $\mathrm N\times\mathrm N$ matrix generally involves $\mathcal O(\mathrm N^4)$ arithmetic operations.  Herein, we present an efficient algorithm that for our particular basis transformation only involves $\mathcal O(\mathrm N^2\log_2\mathrm N)$ operations.  Because this algorithm requires fewer than $\mathcal O(\mathrm N^3)$ operations, for large $\mathrm N$, it could be used as a preprocessing step for quantum computing algorithms for certain applications.  As a demonstration, we apply our algorithm to a Hamiltonian describing a system of relativistic interacting spin-zero bosons and calculate the ground-state energy using the variational quantum eigensolver algorithm on a quantum computer.
	\end{abstract}
	
	\maketitle

	\section{Introduction}

	It has long been recognized that quantum computing offers inherent advantages over classical computing~\cite{Benioff80,Feynman82}, allowing quantum computers to solve certain mathematical tasks otherwise unfeasible~\cite{Shor94,Grover96,Lloyd96_1073}.  Recent advances in fabricating quantum computers with quantum registers containing tens of qubits have stimulated renewed efforts in making practical quantum algorithms to run on these quantum processors, offering polynomial, if not exponential speedup over corresponding classical calculations~\cite{Martinis19_505}.  Such algorithms are now increasingly being used in various application areas, including quantum simulation of quantum chemical systems~\cite{Gambetta17_242,Gambetta19_491,Fischer19,Rubin20}.
	
	One significant challenge in the implementation of existing algorithms, and potentially the development of new ones is the loading of information into the quantum computer~\cite{Aaronson15_291}.  Furthermore, the simulation of Hamiltonian dynamics~\cite{Lloyd96_1073}, the variational quantum eigensolver (VQE) algorithm~\cite{OBrien14_4213}, and the HHL algorithm for linear systems of equations~\cite{Lloyd09_150502} all assume that the linear operator of interest be described in what we herein refer to as the Pauli representation. This representation is a vector representation in a unique basis composed of tensor products of identity and Pauli operators.  Such a basis is particularly amenable to quantum computing, whether for the evaluation of expectation values or the representation of unitary operators used in the quantum logic gates.  While there are efficient methods for certain classes of Hamiltonians~\cite{Ortiz01,Kitaev02_210}, a general transformation that could be applied to arbitrary operators is desirable to expand the reach of these algorithms and create the opportunity for new algorithms.
	
	In this article, we present an algorithm that derives the Pauli coordinates in the Pauli representation of an arbitrary linear operator on a finite-dimensional vector space from its matrix representation.  As a bijective map from one $\mathrm N\times\mathrm N$ matrix to another is described by $\mathrm N^4$ coefficients, an algorithm performing a general transformation of linear operators described by $\mathrm N\times\mathrm N$ matrices would involve at least $\mathcal O(\mathrm N^4)$ arithmetic operations.  As most linear algebra tasks, including eigenvalue decomposition, can be accomplished in $\mathcal O(\mathrm N^3)$ operations, having an $\mathcal O(\mathrm N^4)$ preprocessing step for any subsequent quantum algorithm would render the overall method inefficient.  Herein however, we show that the particular transformation from a matrix representation of a linear operator to the corresponding Pauli representation can be accomplished using an algorithm requiring only $\mathcal O(\mathrm N^2\log_2\mathrm N)$ operations, or less.  This efficient preprocessing algorithm opens the possibility for quantum algorithms of a variety of linear algebra tasks to display an overall quantum advantage.
	
	In the next section, Sec.\,\ref{s.2}, we present an efficient method for generating the sought Pauli coordinates in the Pauli representation of an arbitrary linear operator.  This method is the basis for the algorithm presented and analyzed in Sec.\,\ref{s.3}.  As an illustration, Sec.\,\ref{s.4} presents calculations that produce the ground-state energy for a system of relativistic interacting spin-zero bosons using the transformation algorithm herein as a preprocessing step.  This section covers the derivation of input matrix elements of the Hamiltonian on a restricted Fock space, the transformation itself, and subsequent ground-state energy calculations performed using exact diagonalization and the VQE algorithm with the quantum portion executed on an IBM Q Simulator and the IBM Q Santiago quantum computer.  Lastly, Sec.\,\ref{s.5} summarizes the main conclusions.
	
%=============================================================================%
	\section{Method}
	\label{s.2}
	
	Consider an arbitrary finite linear system and let the state space for this system be the $\mathrm n$-dimensional vector space $\mathcal V_\mathrm n$ over the real or complex field $\mathbb F$.  The system can then be described by a linear operator $\hat A$ in the set of endomorphisms $\operatorname{End}(\mathcal V_\mathrm n)$.  Herein, we define $\hat A$ by its matrix representation
	\begin{equation}
	A=
	\begin{pmatrix}
	a_{0,0} & \hdots & a_{0,j} & \hdots & a_{0,n-1}\\
	\vdots & \ddots & & & \vdots\\
	a_{i,0} & & a_{i,j} & & a_{i,n-1}\\
	\vdots & & & \ddots & \vdots\\
	a_{n-1,0} & \hdots & a_{n-1,j} & \hdots & a_{n-1,n-1}
	\end{pmatrix},
	\label{e.1}
	\end{equation}
	and assume that all matrix elements $a_{i,j}\in\mathbb F$ are given.
	
	Our objective is to represent the linear operator $\hat{\mathrm A}$ on the $\mathrm N$-dimensional Hilbert space $\mathcal H^\mathrm Q$ for a quantum register comprising $\mathrm Q$ identical qubits using the basis composed of tensor products of identity and Pauli operators, where $\mathcal H^\mathrm Q$ denotes the $\mathrm Q^\mathrm{th}$ tensor power of the two-dimensional Hilbert space $\mathcal H$.  To keep track of the qubits in this register, we introduce the index set $\mathcal Q=\{0,1,...,\mathrm Q-1\}$ and label their spaces $\mathcal H_q$, where $\mathcal H_q=\mathcal H$, for all $q\in\mathcal Q$.
	
	Our first task is to inject $\mathcal V_\mathrm n$ into $\mathcal V_\mathrm N\cong\mathcal H^\mathrm Q$, which requires that $\mathrm n\le\mathrm N$, where $\mathrm N=2^\mathrm Q$.  We recommend letting $\mathrm Q=\lceil\log_2\mathrm n\rceil$, but any larger $\mathrm Q$ will work as well.  In any case, we define our injection such that the elements
	\begin{equation}
	\mathrm a_{i,j}=\left\{
	\begin{array}{ll}
	a_{i,j}, & \mathrm{for}~i,j\in\{0,1,...,\mathrm n-1\},\\[5pt]
	\Delta\delta_{i,j}, & \mathrm{for}~i,j\in\{\mathrm n,\mathrm n+1,...,\mathrm N-1\},\\[5pt]
	0, & \mathrm{otherwise},
	\end{array}
	\right.
	\label{e.1b}
	\end{equation}
	form the matrix representation for $\hat{\mathrm A}\in\operatorname{End}(\mathcal V_\mathrm N)$, where $\Delta\in\mathbb F$ is a constant and $\delta$ is the Kronecker delta.  Although $\Delta$ can take any value, including zero, depending on how the final produced representation of $\hat{\mathrm A}$ is going to be used, it can be advantageous to choose $\Delta$ such that its absolute value is large compared to those of all other eigenvalues of $\hat{\mathrm A}$.
	
	Next, we derive an equivalent representation of the linear operator $\hat{\mathrm A}\in\operatorname{End}(\mathcal H^\mathrm Q)$ that takes advantage of the tensor product form of the full transformation monoid
	\begin{equation}
	\operatorname{End}(\mathcal H^\mathrm Q)=\operatorname{End}(\mathcal H_1)\otimes\operatorname{End}(\mathcal H_2)\otimes\ldots\otimes\operatorname{End}(\mathcal H_\mathrm Q).
	\label{e.2}
	\end{equation}
	Consider the operators $\hat{\tau}_0$, $\hat{\tau}_1$, $\hat{\tau}_2$, and $\hat{\tau}_3$ defined by their respective matrix representations
	\begin{equation}
	\arraycolsep=0.04in
	\begin{array}{ccc}
		\tau_0\!=\!\left(\!\begin{array}{cc}
			1 & 0\\ 0 & 0
		\end{array}\!\right)\!,~
		\tau_1\!=\!\left(\!\begin{array}{cc}
			0 & 1\\ 0 & 0
		\end{array}\!\right)\!,~
		\tau_2\!=\!\left(\!\begin{array}{cc}
			0 & 0\\ 1 & 0
		\end{array}\!\right)\!,~
		\tau_3\!=\!\left(\!\begin{array}{cc}
			0 & 0\\ 0 & 1
		\end{array}\!\right)\!.
		\end{array}
	\label{e.3}
	\end{equation}
	It is clear that the set $\{\hat{\tau}_0,\hat{\tau}_1,\hat{\tau}_2,\hat{\tau}_3\}$ forms a basis for $\operatorname{End}(\mathcal H_q)$, for each $q\in\mathcal Q$.  Using the operators in $\{\hat{\tau}_0,\hat{\tau}_1,\hat{\tau}_2,\hat{\tau}_3\}$, we can construct tensor products of the form
	\begin{equation}
	\hat{\mathrm T}_{r_\mathcal Q}=\hat\tau_{r_{\mathrm Q-1}}\otimes\hat\tau_{r_{\mathrm Q-2}}\otimes\ldots\otimes\hat\tau_{r_0},
	\label{e.4}
	\end{equation}
	where $r_\mathcal Q=(r_q)_{q\in\mathcal Q}$ are families of elements in $\mathcal R=\{0,1,2,3\}$ indexed by $\mathcal Q$.  As the set $\{\hat{\mathrm T}_{r_\mathcal Q}\}$ of all such tensor products forms a basis for $\operatorname{End}(\mathcal H^\mathrm Q)$, we have the linear combination
	\begin{equation}
	\hat{\mathrm A}=\sum_{r_\mathcal Q}c^{(0)}_{r_\mathcal Q}\,\hat{\mathrm T}_{r_\mathcal Q},
	\label{e.5}
	\end{equation}
	where the coordinates $c^{(0)}_{r_\mathcal Q}\in\mathbb F$ uniquely specifies our linear operator $\hat{\mathrm A}$.  In other words, there is an isomorphism between the matrix representation $\mathbb F^{\mathrm N\times\mathrm N}$ and the vector representation $\mathbb F^{4^\mathrm Q}$ of $\hat{\mathrm A}$ in the basis $\{\hat{\mathrm T}_{r_\mathcal Q}\}$.  This isomorphism can be described by a one-to-one correspondence between the set of matrix elements $\{\mathrm a_{i,j}\}$ from Eq.\,(\ref{e.1b}) and the coordinate set $\{\mathrm c^{(0)}_{r_\mathcal Q}\}$ that copies the elements
	\begin{equation}
	\mathrm a_{i,j}\mapsto \mathrm c^{(0)}_{r_\mathcal Q}=\mathrm a_{i,j},
	\label{e.6}
	\end{equation}
	for all $i,j\in\mathcal N=\{0,1,...,\mathrm N\}$, in accordance with a separate bijection between the sets $\{(i,j)\}$ and $\{r_\mathcal Q\}$.
	
	Before defining this latter bijection, let us first associate the indices $i$ and $j$ with the families $i_\mathcal Q=(i_q)_{q\in\mathcal Q}$ and $j_\mathcal Q=(j_q)_{q\in\mathcal Q}$, respectively, of elements in $\mathcal B=\{0,1\}$ indexed by $\mathcal Q$.  We generate these families exploiting the isomorphisms $i_\mathcal Q\cong(i)_2$ and $j_\mathcal Q\cong(j)_2$, where $(i)_2$ and $(j)_2$ denote the binary representations of $i$ and $j$, respectively.  Formally, we identify the elements $i_q$ and $j_q$ with the $q^\mathrm{th}$ digits in $(i)_2$ and $(j)_2$, respectively, for all $q\in\mathcal Q$.  Lastly, for each generated pair of families $i_\mathcal Q$ and $j_\mathcal Q$, we produce the family $r_\mathcal Q$ using the mapping $\mathcal B\times\mathcal B\rightarrow\mathcal R$ defined by
	\begin{equation}
	(i_q,j_q)\mapsto r_q=2*i_q+j_q,
	\label{e.7}	
	\end{equation}
	again for all $q\in\mathcal Q$.  See Table~\ref{t.1} below.  With a well-defined bijection from $\{(i,j)\}$ to $\{r_\mathcal Q\}$, we can now obtain all the coordinates $\mathrm c^{(0)}_{r_\mathcal Q}$ in Eq.\,(\ref{e.5}) using Eq.\,(\ref{e.6}).
			
	Equipped with a vector representation of the linear operator $\hat{\mathrm A}\in\operatorname{End}(\mathcal H^\mathrm Q)$, we next formulate a basis transformation of $\operatorname{End}(\mathcal H_q)$ to the basis composed of operators in $\{\hat{\mathrm I},\hat{\sigma}_x,\hat{\sigma}_y,\hat{\sigma}_z\}$, where $\hat{\mathrm I}$ is the two-dimensional identity operator and $\hat{\sigma}_x$, $\hat{\sigma}_y$, and $\hat{\sigma}_z$ are the three Pauli operators.  However, rather than using the basis $\{\hat{\mathrm I},\hat{\sigma}_x,\hat{\sigma}_y,\hat{\sigma}_z\}$ for $\mathcal H$ directly in the construction of tensor products for $\mathcal H^\mathrm Q$, let us first introduce the closely related basis $\{\hat{\sigma}_0,\hat{\sigma}_1,\hat{\sigma}_2,\hat{\sigma}_3\}$ defined by the operators $\hat{\sigma}_0=\hat{\mathrm I}$, $\hat{\sigma}_1=\hat{\sigma}_x$, $\hat{\sigma}_2=-i\hat{\sigma}_y$, and $\hat{\sigma}_3=\hat{\sigma}_z$, where $i$ is the imaginary unit.  This basis has the advantage that all nonzero elements of the matrix representations of its basis operators,
	\begin{equation}
	\arraycolsep=0.04in
	\begin{array}{ccc}
		\sigma_0\!=\!\left(\!\begin{array}{cc}
			1 & 0\\ 0 & 1
		\end{array}\!\right)\!,\,
		\sigma_1\!=\!\left(\!\begin{array}{cc}
			0 & 1\\ 1 & 0
		\end{array}\!\right)\!,\,
		\sigma_2\!=\!\left(\!\begin{array}{cc}
			0 & \!-1\\ 1 & 0
		\end{array}\!\!\right)\!,\,
		\sigma_3\!=\!\left(\!\begin{array}{cc}
			1 & 0\\ 0 & \!-1
		\end{array}\!\!\right)\!,
		\end{array}
	\label{e.8}
	\end{equation}
	respectively, are contained in $\mathcal S=\{\pm1\}$, which in some instances might result in a faster implementation of the algorithm below.

	The tensor products of the operators in $\{\hat{\mathrm I},\hat{\sigma}_x,\hat{\sigma}_y,\hat{\sigma}_z\}$ can then be expressed on the form
	\begin{equation}
	\hat{\mathrm S}_{r_\mathcal Q}=\Theta_{r_\mathcal Q}\,\hat\sigma_{r_{\mathrm Q-1}}\otimes\hat\sigma_{r_{\mathrm Q-2}}\otimes\ldots\otimes\hat\sigma_{r_0},
	\label{e.9}
	\end{equation}
	where the phase factor
	\begin{equation}
	\Theta_{r_\mathcal Q}=i\,^{\sum_q\!\delta_{r_q,2}}
	\label{e.10}
	\end{equation}
	ensures that we form the desired basis $\{\hat{\mathrm S}_{r_\mathcal Q}\}$ for $\operatorname{End}(\mathcal H^\mathrm Q)$.  In this basis, the operator $\hat{\mathrm A}\in\operatorname{End}(\mathcal H^\mathrm Q)$ is given by the linear combination
	\begin{equation}
	\hat{\mathrm A}=\sum_{r_\mathcal Q}\mathrm c_{r_\mathcal Q}\,\hat{\mathrm S}_{r_\mathcal Q},
	\label{e.11}
	\end{equation}
	where $\mathrm c_{r_\mathcal Q}\in\mathbb F$ are the $4^\mathrm Q$ Pauli coordinates we seek.
	
	The basis transformation from $\{\hat{\mathrm T}_{r_\mathcal Q}\}$ to $\{\hat{\mathrm S}_{r_\mathcal Q}\}$ is given by
	\begin{equation}
	\hat{\mathrm S}_{r_\mathcal Q}=\Theta_{r_\mathcal Q}\sum_{r_\mathcal Q'}\mathrm M_{r_\mathcal Q,r_\mathcal Q'}\,\hat{\mathrm T}_{r_\mathcal Q'},
	\label{e.12}
	\end{equation}
	for all families $r_\mathcal Q\in\mathcal R^\mathrm Q$, where $\mathrm M_{r_\mathcal Q,r_\mathcal Q'}$ are $4^\mathrm Q\times4^\mathrm Q=\mathrm N^4$ coefficients describing the basis transformation operator $\widehat{\mathrm M}\in\operatorname{End}(\operatorname{End}(\mathcal H^\mathrm Q))$.  The transformation operator is of the form
	\begin{equation}
	\widehat{\mathrm M}=\bigotimes_q\widehat{\mathrm m}_q,
	\label{e.13}
	\end{equation}
	and it follows from the definitions in Eqs.\,(\ref{e.3}) and (\ref{e.8}) that the matrix representations of the operators $\widehat{\mathrm m}_q$ are identical for all $q\in\mathcal Q$ with $\mathrm m_q=\mathrm m$, where
	\begin{equation}
	\mathrm m=
	\begin{pmatrix}
	1 & 0 & 0 & 1\\
	0 & 1 & 1 & 0\\
	0 & 1 & -1 & 0\\
	1 & 0 & 0 & -1
	\end{pmatrix}.
	\label{e.14}
	\end{equation}
	This transformation matrix is involutory up to a structure constant, satisfying $\mathrm m^2=2\mathrm I$, where $\mathrm I$ is a unit matrix.  Thus, the corresponding inverse transformation matrix is given by $\mathrm m^{-1}=\mathrm m/2$.
		
	Putting everything together, the sought coordinates can now be obtained from the bijective mapping $\mathbb F^{4^\mathrm Q}\rightarrow\mathbb F^{4^\mathrm Q}$ defined by $\mathrm c^{(0)}_{r_\mathcal Q}\mapsto \mathrm c_{r_\mathcal Q}$, for all $r_\mathcal Q\in\mathcal R^\mathrm Q$, where we have
	\begin{equation}
	\mathrm c_{r_\mathcal Q} = 2^{-\mathrm Q}\,\Theta_{r_\mathcal Q}^{-1}\sum_{r_\mathcal Q'}\prod_{q\in\mathcal Q}\mathrm m_{r_q,r_q'}\,\mathrm c^{(0)}_{r_\mathcal Q'},
	\label{e.15}
	\end{equation}
	from Eqs.\,(\ref{e.4}--\ref{e.5},\,\ref{e.9},\,\ref{e.11}--\ref{e.13}) and the relation $\mathrm m^{-1}=\mathrm m/2$.  Because the basis transformation in Eq.\,(\ref{e.12}) is separable, as Eq.\,(\ref{e.13}) shows, we can perform the transformation in $\mathrm Q$ independent steps.  Specifically, we have
	\begin{equation}
	\mathrm c_{r_\mathcal Q} = 2^{-\mathrm Q}\,\Theta_{r_\mathcal Q}^{-1}\,\mathrm c^{(\mathrm Q)}_{r_\mathcal Q},
	\label{e.16}
	\end{equation}
	where $\mathrm c^{(\mathrm Q)}_{r_\mathcal Q}$ is obtained after iterating
	\begin{equation}
	\mathrm c^{(q+1)}_{r_\mathcal Q} = \prod_{q'\ne q}\delta_{r_{q'},r_{q'}'}\sum_{r_q'}\mathrm m_{r_q,r_q'}\,\mathrm c^{(q)}_{r_\mathcal Q'},
	\label{e.17}
	\end{equation}
	for all $r_\mathcal Q\in\mathcal R^\mathrm Q$, over all $q\in\mathcal Q$.	 The presence of the delta functions is a manifestation of the separability of the transformation.  Note that it follows from Eq.\,(\ref{e.14}) that for our particular basis transformation, there are exactly two nonzero terms in the sum in Eq.\,(\ref{e.17}).  Thus, we can express this equation as the linear combination
	\begin{equation}
	\mathrm c^{(q+1)}_{r_\mathcal Q} = \theta_{r_q}\,\mathrm c^{(q)}_{r_\mathcal Q}+\theta_{r_q^*}\,\mathrm c^{(q)}_{r_\mathcal Q^*},
	\label{e.18}
	\end{equation}
	for all $q\in\mathcal Q$, where $r_\mathcal Q^*\in\mathcal R^\mathrm Q$ is defined by
	\begin{equation}
	r_{q'}^*=\left\{
	\begin{array}{ll}
	3-r_{q'}, & \mathrm{for}~q'=q,\\[5pt]
	r_{q'}, & \mathrm{for}~q'\ne q,
	\end{array}
	\right.
	\label{e.18b}
	\end{equation}
	for all $q'\in\mathcal Q$, and where the two coefficients $\theta_{r_q},\theta_{r_q^*}\in\mathcal S$ follows from Eq.\,(\ref{e.14}) and are given in Table~\ref{t.1}.  With these definitions, we can now obtain all the $4^\mathrm Q$ Pauli coordinates $\mathrm c_{r_\mathcal Q}$ using Eqs.\,(\ref{e.10},\,\ref{e.16},\,\ref{e.18}).
\begin{table}[h]
	\centering
	\caption{Isomorphism between the $q^\mathrm{th}$ elements in the families $i_\mathcal Q$, $j_\mathcal Q$, and $r_\mathcal Q$, describing the positions in the matrix and vector representations of the linear operator $\hat{\mathrm A}\in\operatorname{End}(\mathcal H^\mathrm Q)$.  The coefficients $\theta_{r_q}$ and $\theta_{r_q^*}$ describe the coordinate relationship in Eq.\,(\ref{e.18}).}
	\begin{tabular}{c|c|c||c|c|c}
		\hline
		$\quad(i_q,j_q)\quad$ & $\quad r_q\quad$  & $\quad\theta_{r_q}\quad$ & $\quad(i_q^*,j_q^*)\quad$  & $\quad r_q^*\quad$  & $\quad\theta_{r_q^*}\quad$\\
		\hline\hline
		$(0,0)$ & $0$ & $+1$ & $(1,1)$ & $3$ & $+1$\\
		$(0,1)$ & $1$ & $+1$ & $(1,0)$ & $2$ & $+1$\\
		$(1,0)$ & $2$ & $-1$ & $(0,1)$ & $1$ & $+1$\\
		$(1,1)$ & $3$ & $-1$ & $(0,0)$ & $0$ & $+1$\\
		\hline
	\end{tabular}
	\label{t.1}
\end{table}
	
%=============================================================================%
	\section{Algorithm}
	\label{s.3}
	
	\subsection{General matrix representations}

	To implement the method described above, we use the following algorithm:
	\begin{enumerate}
		\item Identify the $\mathrm n\times\mathrm n$ matrix representing the linear operator $\hat A\in\operatorname{End}(\mathcal V_\mathrm n)$ over the real or complex field $\mathbb F$.
		\item Using Eq.\,(\ref{e.1b}), define the matrix elements $\mathrm a_{i,j}$ describing $\hat{\mathrm A}\in\operatorname{End}(\mathcal V_\mathrm N)$, indexed by $\mathcal N\times\mathcal N$, by the matrix elements of $\hat A$ and copies of the free parameter $\Delta\in\mathbb F$.
		\item Associate each matrix index pair $(i,j)$ with a vector index $r_\mathcal Q$ using Eq.\,(\ref{e.7}), for all $q\in\mathcal Q$, where $i_q$ and $j_q$ are the $q^\mathrm{th}$ digits in binary representations of $i$ and $j$, respectively.
		\item Using Eq.\,(\ref{e.6}), map the elements $\mathrm a_{i,j}$ to the initial coordinates $\mathrm c^{(0)}_{r_\mathcal Q}$ describing $\hat{\mathrm A}\in\operatorname{End}(\mathcal H^\mathrm Q)$ in the basis $\{\hat{\mathrm T}_{r_\mathcal Q}\}$.
		\item Iterate over $q\in\mathcal Q$, and for each such iteration calculate a new coordinate set $\{\mathrm c^{(q+1)}_{r_\mathcal Q}\}$ using Eq.\,(\ref{e.18}), where $r_\mathcal Q^*$ and the signs $\theta_{r_q},\theta_{r_q^*}\in\mathcal S$ are defined by Eq.\,(\ref{e.18b}) and Table~\ref{t.1}, respectively.
		\item Generate the final coordinates $\mathrm c_{r_\mathcal Q}$ describing $\hat{\mathrm A}\in\operatorname{End}(\mathcal H^\mathrm Q)$ in the desired basis $\{\hat{\mathrm S}_{r_\mathcal Q}\}$ using Eq.\,(\ref{e.16}) with $\Theta_{r_\mathcal Q}$ given by Eq.\,(\ref{e.10}).
	\end{enumerate}
	
	As the transformations in step 5 requires $\mathrm Q$, or equivalently $\log_2\mathrm N$ iterations of Eq.\,(\ref{e.18}), each involving the calculation of $\mathrm N^2$ coordinates, the total number of coordinates that need to be calculated is
	\begin{equation}
	\mathrm L_{\mathrm{dense}} = \mathrm N^2\log_2\mathrm N.
	\label{e.21}
	\end{equation}
	From this result, we conclude that the number of needed arithmetic operations in a fast implementation scales as $\mathcal O(\mathrm N^2\log_2\mathrm N)$.
	
	The algorithm herein is thus remarkably fast, considering that the $\mathrm N^2\times\mathrm N^2$ matrix $\mathrm M$ describing the basis transformation operator $\widehat{\mathrm M}$ in Eq.\,(\ref{e.13}) has $\mathrm N^4$ elements and ordinarily would require at least $\mathcal O(\mathrm N^4)$ operations.
	
	\begin{figure}[t]
		\includegraphics[width=\columnwidth]{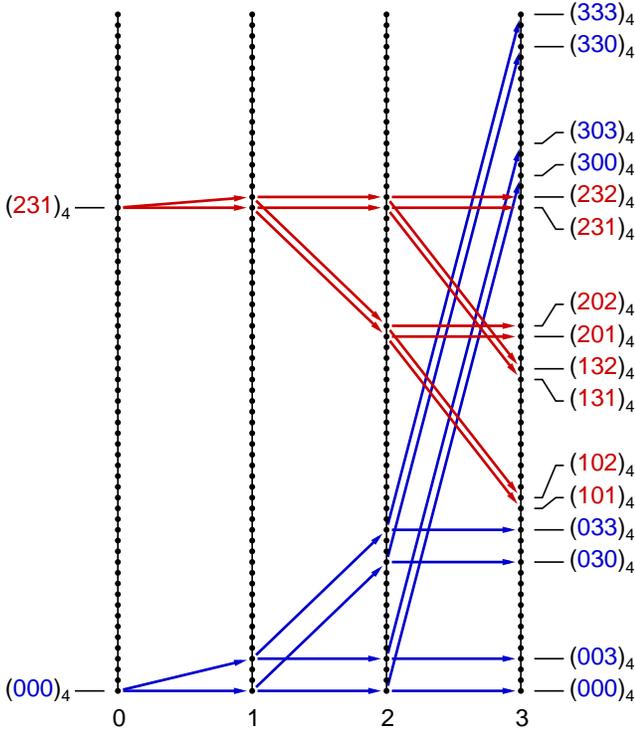}
		\caption{Coordinate couplings for $\mathrm Q=3$.  In each iteration $q\in\mathcal Q$, each of the $\mathrm N^2=64$ old coordinates $c^{(q)}_{r_\mathcal Q}$ with index $r_\mathcal Q$ (depicted by black dots) contributes to two new coordinates $\mathrm c^{(q+1)}_{r_\mathcal Q}$ and $\mathrm c^{(q+1)}_{r_\mathcal Q^*}$, where $r_\mathcal Q^*$ is given by Eq.\,(\ref{e.18b}).  The matrix representation of the full transformation operator therefore contains $\mathrm N^3$ nonzero elements and cannot be calculated explicitly in our $\mathcal O(\mathrm N^2\log_2\mathrm N)$ algorithm.}
		\label{f.1}
	\end{figure}
	It is also worth considering what would happen if we were to evaluate Eq.\,(\ref{e.15}) directly without taking advantage of the separability of the transformation.  As each coordinate in Eq.\,(\ref{e.18}) is a linear combination of two previously obtained coordinates, one finds that after $\mathrm Q$ iterations, each final coordinate $\mathrm c_{r_\mathcal Q}$ is a linear combination of $2^\mathrm Q$ distinct initial coordinates $\mathrm c^{(0)}_{r_\mathcal Q}$.  Conversely, because of the transformation is involutory, each initial coordinate contributes to exactly $2^\mathrm Q$ final coordinates.  See Fig.\,\ref{f.1}.  This inverse relationship is useful to track coordinates when the linear operator $\hat A$ is represented by a sparse matrix.
	
	In the general case, there are exactly $\mathrm N$ nonzero terms in the sum in Eq.\,(\ref{e.15}) and a direct evaluation of the $\mathrm N^2$ coordinates using Eq.\,(\ref{e.15}) would require at least $\mathcal O(\mathrm N^3)$ operations.  Remarkably, we could not even compute explicitly, let alone store the $\mathrm N^3$ nonzero matrix elements describing the transformation $\widehat{\mathrm M}$, while retaining the $\mathcal O(\mathrm N^2\log_2\mathrm N)$ scaling of our algorithm.
	
	\begin{figure}[t]
		\centering
		\includegraphics[width=\columnwidth]{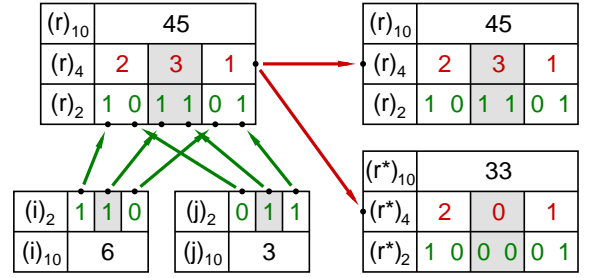}
		\caption{Index mappings and couplings in different base representations.  Owing to the qubit-wise mapping in Eq.\,(\ref{e.7}), the binary representations $(i)_2$ and $(j)_2$ of the matrix indices $i\cong(i)_{10}$ and $j\cong(j)_{10}$, respectively, are interlaced in $(r)_2$ representing the coordinate index $r_\mathcal Q$.  Moreover, the coordinate with index $r_\mathcal Q$ couples to itself and the coordinate with index $r_\mathcal Q^*$ given by Eq.\,(\ref{e.18b}).  The mapping $r_q\mapsto r_q^*=3-r_q$ can be executed by flipping the two bits in the $q^\mathrm{th}$ crumb (gray) of the binary representation of $r_\mathcal Q$ to generate $r_\mathcal Q^*$.}
		\label{f.2}
	\end{figure}
	Owing to the digit-wise mapping in Eq.\,(\ref{e.7}), each coordinate index element $r_q\in\mathcal R$ can be stored in a unit composed of exactly two bits---also known as a crumb.  Collectively, the most significant bits in each crumb form a binary representation of the row index $i$, and the least significant bits form a binary representation of the column index $j$.  Thus, we obtain $r_\mathcal Q$ for each matrix element index pair $(i,j)$ by ``interlacing'' $i$ and $j$.  For instance, the element at row six, column three with $i\cong(6)_{10}\cong(110)_2$ and $j\cong(3)_{10}\cong(011)_2$ immediately maps to the coordinate at $r_\mathcal Q\cong(101101)_2\cong(231)_4$, as shown in the left-hand side of Fig.~\ref{f.2}.
	
	Other operations can also easily be implemented in a bit representation.  First, we generate the coordinate index $r_\mathcal Q^*$ by copying $r_\mathcal Q$ and flipping the two bits in the $q^\mathrm{th}$ crumb.  Second, we set the sign $\theta_{r_q}$ to $+1$ ($-1$), when the $(2q)^\mathrm{th}$ bit is $0$ ($1$).  We always have $\theta_{r_q^*}=+1$.  Lastly, we calculate the exponent in $\Theta_{r_\mathcal Q}$ by counting the number of crumbs that equal $(10)_2$.
	
	For parallel implementations of the algorithm, it is useful to note that one can take advantage of the pairing of all the coordinates in $\{\mathrm c^{(q)}_{r_\mathcal Q}\}$.  If all $\mathrm c^{(q)}_{r_\mathcal Q}$ and $\mathrm c^{(q)}_{r_\mathcal Q^*}$ are kept together in the distribution of the coordinate array, one can update each coordinate pair using
	\begin{equation}
	\begin{pmatrix}
	\mathrm c^{(q+1)}_{r_\mathcal Q}\\
	\mathrm c^{(q+1)}_{r_\mathcal Q^*}
	\end{pmatrix}
	=
	\begin{pmatrix}
	(-1)^{i_q} & 1\\
	1 & (-1)^{i_q^*}
	\end{pmatrix}
	\begin{pmatrix}
	\mathrm c^{(q)}_{r_\mathcal Q}\\
	\mathrm c^{(q)}_{r_\mathcal Q^*}
	\end{pmatrix}.
	\label{e.new}
	\end{equation}
	This allows the coordinate data to be stored in a single $\mathrm N^2$ array and be locally updated.  Efficient data storage is important, as we have found that our $\mathrm N^2\log_2\mathrm N$ algorithm is fast enough in our parallel implementation that data storage, which scales as $\mathrm N^2$, is the real limiting factor.
	
	\subsection{Sparse matrix representations}

	The algorithm for sparse matrix representations of $\hat A$ is the same as that for the general case above, except that we track and only operate on nonzero coordinates in steps 4--6.

	Let the number of initial nonzero coordinates be $l$.  In the sparse limit, nonzero coordinates do not couple to each other within our transformation.  The number of nonzero coordinates then doubles with each iteration, so that the number of coordinates after iteration $q$ is $l_q=2^{q+1}l$.  When the number of initial nonzero coordinates exceeds $\mathrm N$, at most $\mathrm Q^*$ iterations can be performed before $2^{q+1}l$ exceeds the maximum $\mathrm N^2$ coordinates.  Assuming that the coordinate doubling continues up to this point, we have $\mathrm Q^*=\min\{\lfloor-\log_2\ell\rfloor,\mathrm Q\}$, where $\ell=l/\mathrm N^2$ is the initial coordinate density.  The produced coordinate density during iteration $q$ can then be expressed as
	\begin{equation}
	\ell_q=\left\{
	\begin{array}{ll}
	2^{q+1}\ell, & \mathrm{when}~q\le \mathrm Q^*-1,\\[5pt]
	1, & \mathrm{when}~q>\mathrm Q^*-1,
	\end{array}
	\right.
	\label{e.22}
	\end{equation}
for all $q\in\mathcal Q$.  This represents the worst case scenario, as in reality, some nonzero coordinates could couple to each other, which would lead to a lower density, in particular when $q\lesssim\mathrm Q^*$.

	To determine how the number of operations scales at large $\mathrm N$, we estimate the total number of nonzero coordinates $\mathrm L$ that needs to be calculated.  In the worst case scenario mentioned above, the total number of nonzero coordinates produced in all $\mathrm Q$ iterations is
	\begin{equation}
	\mathrm L = \mathrm N^2\sum_{q\in\mathcal Q}\ell_q,
	\label{e.23}
	\end{equation}
	where the coordinate density $\ell_q$ is given in Eq.\,(\ref{e.22}).
	
	\begin{figure}
		\centering
		\includegraphics{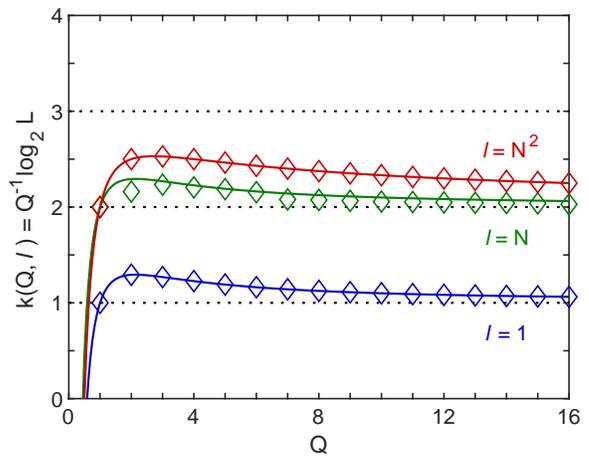}
		\caption{Scaling of the number of nonzero coordinates in our algorithm.  The diamond data points, representing the actual number of calculated coordinates for random initial coordinate sets, match exactly the values from the corresponding curves obtained from Eqs.\,(\ref{e.24}) and (\ref{e.21}) for $l=1$ and $l=\mathrm N^2$, respectively.  In the case $l=\mathrm N$, the numerical data points are all below the corresponding curve obtained from either Eqs.\,(\ref{e.24}) or (\ref{e.25}), owing to nonzero coordinate couplings.  The dashed lines represent linear ($\mathrm k=1$), quadratic ($\mathrm k=2$), and cubic ($\mathrm k=3$) scaling for $\mathrm N$ large.  Even in the worst case scenario, our algorithm scaling is well below the classical cubic scaling for conventional linear algebra routines, as well as any implementation using the full transformation matrix ($\mathrm k=4$).}
		\label{f.3}
	\end{figure}
	The expression for the total number of coordinates takes two different forms depending on whether $\mathrm Q^*$ is equal to or less than $\mathrm Q$.  The critical point is when the initial number of nonzero coordinates is $l=\mathrm N$---or equivalently $\ell=1/\mathrm N$.  For sparse matrix representations with $l\le\mathrm N$, we have $\mathrm Q^*=\mathrm Q$, for which the upper bound for the total number of nonzero coordinates that we need to calculate is
	\begin{equation}
	\mathrm L_{\mathrm{sparse}} = 2(\mathrm N-1)l.
	\label{e.24}
	\end{equation}
	This expression shows that in the very sparse region, where $l$ is a fixed number independent of $\mathrm N$, any fast implementation of our algorithm scales linearly with $\mathrm N$.  Evidence of this scaling can also be seen in the blue curve in Fig.\,\ref{f.3}, representing $l=1$.  Specifically, we note that this curve for large $\mathrm Q$ approaches $\mathrm k=1$, which corresponds to $\mathcal O(\mathrm N)$ scaling.

	In the case $l$ is a fraction of $\mathrm N$, the scaling becomes quadratic.  This scaling is also immediately evident from Eq.\,(\ref{e.24}).  This case is illustrated by the green curve in Fig.\,\ref{f.3}, representing $l=\mathrm N$, which as expected approach $\mathrm k=2$ asymptotically, for large $\mathrm Q$, which corresponds to $\mathcal O(\mathrm N^2)$ scaling.

	When the initial number of nonzero coordinates $l\ge\mathrm N$, the number of coordinates becomes saturated.  The sum in Eq.\,(\ref{e.23}) then splits into two types of terms.  Using $2^{\mathrm Q^*}\ell\approx1$, we arrive at the approximate expression
	\begin{equation}
	\mathrm L_{\mathrm{intermediate}} = \mathrm N^2\left[\log_2\mathrm N\ell+2(1-\ell)\right].
	\label{e.25}
	\end{equation}
	For $\ell=1/\mathrm N$, the first term is zero and the right-hand side coincides with that in Eq.\,(\ref{e.24}) with $l=\mathrm N$.  Thus, at this critical point, we have $\mathrm L_{\mathrm{intermediate}}=\mathrm L_{\mathrm{sparse}}=2\mathrm N(\mathrm N-1)$, which as already mentioned scales quadratically.
	
	Increasing the initial nonzero coordinate density $\ell$ eventually lands us at the dense limit $\ell=1$.  At this point, we have $\mathrm L_{\mathrm{intermediate}}=\mathrm L_{\mathrm{dense}}=\mathrm N^2\log_2\mathrm N$ as expected.  The upper bound for the number of nonzero coordinates that we need to calculate is shown by the red curve representing $l=\mathrm N^2$ in Fig.\,\ref{f.3}.  Because the difference between $\mathcal O(\mathrm N^2)$ and $\mathcal O(\mathrm N^2\log_2\mathrm N)$ is relatively small when $\mathrm N$ is not too large, one needs to make a judgement whether the overhead of tracking nonzero coordinates is worthwhile on a case by case basis.  In any case, while not quite quadratic, the central point of this work is that this curve is well below the well known $\mathrm k=3$ scaling for conventional linear algebra routines.

	To see how our total-number-of-nonzero-coordinate expressions above hold up in practice, we have performed coordinate tracking within our algorithm and applied our implementation to initial coordinate sets with $l$ nonzeros at random indices $r_\mathcal Q$.  The results are shown as diamonds in Fig.~\ref{f.3} for $l=1, \mathrm N, \mathrm N^2$.  In the first case ($l=1$) when there is only a single nonzero element, no nonzero coordinate couplings are possible, and thus the calculated $\mathrm L$ is a lower limit that matches exactly that given by Eq.\,(\ref{e.24}) for $l=1$.  In the third case ($l=\mathrm N^2$), the initial coordinate set is dense with $\ell=1$, which means that all coordinates are generally nonzero.  This is the upper limit, which also does not permit variability.  Consequently, the calculated $\mathrm L$ is exactly that of Eq.\,(\ref{e.21}).  In between these two limits, we observe variability in the calculated $\mathrm L$ caused by nonzero coordinate couplings.  See how the data points do not exactly match the curve for $l=\mathrm N$.  Also note that all data points are below the curve, which is expected as the curve as mentioned above, is at this critical point not only given by the approximate Eq.\,(\ref{e.25}) but also the upper bound in Eq.\,(\ref{e.24}) for  $l=\mathrm N$.
	
	This observation confirms that a finite number of coordinate couplings only reduce the amount of needed calculations and never increase it.  The most extreme case of this effect occurs, for instance, when the initial coordinate set represents a diagonal operator.  In this case, $\ell_q=\mathrm N^{-1}$, for all $q\in\mathcal Q$, resulting in $\mathrm L=\mathrm N\log_2\mathrm N$ instead of $\mathrm L=2\mathrm N(\mathrm N-1)$.

%=============================================================================%
	\section{Simulating relativistic interacting spin-zero bosons}
	\label{s.4}

	The Jordan-Wigner transform~\cite{Jordan28} allows for the efficient mapping of a Hamiltonian of a fermionic system to a Hamiltonian of a quantum register by representing the fermionic creation and annihilation operators as tensor products of identity and Pauli operators~\cite{Ortiz01}.  However, no corresponding transformation exists for the bosonic operators that satisfy the bosonic commutation relations.  As a result, an alternative method is needed to transform bosonic systems.
	
	To illustrate how the method herein could be used, we solve below the ground-state energy of an interacting system of relativistic spin-zero bosons.  As is customary in quantum field theory, we describe this system by a Hamiltonian
	\begin{equation}
	\hat H = \hat H_0 + \hat V
	\end{equation}
	that separates the terms $\hat H_0$ describing the corresponding free non-interacting system and the terms $\hat V$ describing the boson interactions.  Free relativistic spin-zero bosons are described by the Klein--Gordon Hamiltonian
	\begin{equation}
	\hat H_0 =\frac{1}{2}\int\Big\{\pi^2(\vec r)+\big[\hbar c\nabla\phi(\vec r)\big]^2+\big[mc^2\phi(\vec r)\big]^2\Big\}\,\mathrm d^3r,
	\end{equation}
	where $\hbar$, $c$, and $m$ are the reduced Planck constant, the speed of light, and the boson mass, respectively, and $\phi(\vec r)$ and $\pi(\vec r)$ are quantum fields satisfying the commutation relations $[\phi(\vec r),\pi(\vec r')]=i\delta(\vec r-\vec r')$ and $[\phi(\vec r),\phi(\vec r')]=[\pi(\vec r),\pi(\vec r')]=0$.  The interacting terms are herein modeled by the quartic interaction
	\begin{equation}
	\hat V =  \frac{\lambda}{4!}\big(\hbar c\big)^3\int\phi^4(\vec r)\,\mathrm d^3r,
	\end{equation}
	where $\lambda$ is a dimensionless coupling constant.
	
	We assume that the bosons are contained in a three-dimensional box with Cartesian dimensions $L_x\!\times\!L_y\!\times\!L_z$ centered at the Cartesian coordinate $(L_x,L_y,L_z)/2$.  Moreover, we assume that this box has $M=M_x\!\times\!M_y\!\times\!M_z$ modes described by the wave vectors $\vec k_{\vec \mu}=(k_{\mu_x},k_{\mu_y},k_{\mu_z})$, where $\vec\mu=(\mu_x,\mu_y,\mu_z)$ and $k_{\mu_\alpha}L_\alpha=\mu_\alpha\pi$, for all integers $\mu_\alpha\in\mathcal M_\alpha=\{1,2,...,M_\alpha\}$, for all Cartesian components $\alpha\in\{x,y,z\}$.

	With these modes, we can define the fields
	\begin{align}
	\phi(\vec r) &= \sum_{\vec \mu}\frac{1}{\sqrt{2\hbar\omega_{\vec \mu}}}\left(\hat a_{\vec \mu}+\hat a_{\vec \mu}^\dagger\right)\prod_{\alpha}\sqrt{\frac{2}{L_\alpha}}\sin k_{\mu_\alpha}\!r_\alpha,\\
	\pi(\vec r) &= \sum_{\vec \mu}(-i)\sqrt{\frac{\hbar\omega_{\vec \mu}}{2}}\left(\hat a_{\vec \mu}-\hat a_{\vec \mu}^\dagger\right)\prod_{\alpha}\sqrt{\frac{2}{L_\alpha}}\sin k_{\mu_\alpha}\!r_\alpha,
	\end{align}
	where $\vec r=(r_x,r_y,r_z)$ with $r_\alpha=\alpha$, for all components $\alpha$, and $\hat a_{\vec \mu}$ and $\hat a_{\vec \mu}^\dagger$ are annihilation and creation operators for bosons in mode $\vec \mu$ with energy
	\begin{equation}
	\hbar\omega_{\vec \mu} = \sqrt{\big|\hbar c\vec k_{\vec \mu}\big|^2 + \big(mc^2\big)^2}.
	\label{e.32}
	\end{equation}
	Using the bosonic commutation relations $[\hat a_{\vec \mu},\hat a_{\vec \nu}^\dagger]=\delta_{\vec\mu,\vec\nu}$ and $[\hat a_{\vec \mu},\hat a_{\vec \nu}]=[\hat a_{\vec \mu}^\dagger,\hat a_{\vec \nu}^\dagger]=0$, for all $\vec \mu,\vec \nu\in\mathcal M_x\times\,\mathcal M_y\times\,\mathcal M_z$, we find as expected that this choice of fields diagonalizes the Hamiltonian of the free non-interacting system, which becomes
	\begin{equation}
	\hat H_0 = \sum_{\vec \mu}\hbar\omega_{\vec \mu}\,\hat{a}_{\vec \mu}^\dagger\hat{a}_{\vec \mu},
	\end{equation}
	after the the zero-point energy has been dropped.  The energy of the $n_{\vec\mu}$ particles in mode $\vec\mu$ of the non-interacting system is then $E_{\vec\mu}=\hbar\omega_{\vec \mu}n_{\vec \mu}$.  As neither the mode frequencies $\hbar\omega_{\vec \mu}$ nor the particle numbers $n_{\vec \mu}$ can be negative, the vacuum state $|0\rangle$, for which $n_{\vec \mu}=0$, for all $\mu$, is the ground state of the non-interacting system with energy $E_0=0$.
	
	In this representation, the Hamiltonian describing the boson interactions becomes
	\begin{align}
	\hat V = \sum_{\vec\mu\vec\nu\vec\xi\vec o} V_{\vec\mu\vec\nu\vec\xi\vec o}\Big(&3\delta_{\vec\mu\vec\nu}\delta_{\vec\xi\vec o}+6\delta_{\vec\mu\vec\nu}\hat a_{\vec\xi}\hat a_{\vec o}+12\delta_{\vec\mu\vec\nu}\hat a^\dagger_{\vec\xi}\hat a_{\vec o}\nonumber\\
	&\!+6\delta_{\vec\mu\vec\nu}\hat a^\dagger_{\vec\xi}\hat a^\dagger_{\vec o}+\hat a_{\vec\mu}\hat a_{\vec\nu}\hat a_{\vec\xi}\hat a_{\vec o}+4\hat a^\dagger_{\vec\mu}\hat a_{\vec\nu}\hat a_{\vec\xi}\hat a_{\vec o}\nonumber\\[8pt]
	&\!+6\hat a^\dagger_{\vec\mu}\hat a^\dagger_{\vec\nu}\hat a_{\vec\xi}\hat a_{\vec o}+4\hat a^\dagger_{\vec\mu}\hat a^\dagger_{\vec\nu}\hat a^\dagger_{\vec\xi}\hat a_{\vec o}+\hat a^\dagger_{\vec\mu}\hat a^\dagger_{\vec\nu}\hat a^\dagger_{\vec\xi}\hat a^\dagger_{\vec o}\Big),
	\end{align}
	with the coefficients
	\begin{align}
	V_{\vec\mu\vec\nu\vec\xi\vec o}&=\frac{\lambda}{4!}\frac{1}{\Omega}\bigg(\frac{\hbar c}{4}\bigg)^3\frac{1}{\big(2\hbar\big)^2\!\!\sqrt{\vphantom{\hbar}\omega_{\vec\mu}\,\omega_{\vec\nu}\,\omega_{\vec\xi}\,\omega_{\vec o}}}\nonumber\\
	\times&\prod_\alpha\!\sum_{s_\mu s_\nu s_\xi s_o}\!\!\!(-1)^{s_\mu+s_\nu+s_\xi+s_o}\delta_{s_\mu\mu_\alpha+s_\nu\nu_\alpha+s_\xi\xi_\alpha+s_o o_\alpha,0},
	\end{align}
	where $\Omega=L_xL_yL_z$ is the volume of the box and the sum is over all signs $s_\mu,s_\nu,s_\xi,s_o\in\mathcal S$.  Note that the Kronecker delta provides wave vector conservation for each Cartesian component $\alpha$.
	
	Because any number of bosons can occupy a single mode, we must limit the number of particles, as well as the number of modes in the box.  Denote the maximum number of particles $N$.  We can then form a basis from all possible Fock states
	\begin{equation}
	|n_1,n_2,...,n_M\rangle = \prod_{\vec\mu}\frac{1}{\sqrt{n_{\vec\mu}\,!}}\big(a^\dagger_{\vec\mu}\big)^{n_{\vec\mu}}|0\rangle,
	\end{equation}
	where the total number of particles $\sum_{\vec\mu}n_{\vec\mu}\le N$.  The number of such Fock states, and hence the dimension of our Hilbert space is
	\begin{equation}
	\mathrm n =
	\frac{(M+N)!}{M!\,N!}.
	\label{e.37}
	\end{equation}
	Letting the arbitrary linear operator $\hat A$ be the Hamiltonian $\hat H$, the matrix elements describing our operator are given by
	\begin{equation}
	a_{i,j}=\langle n_1^{(i)},n_2^{(i)},...,n_M^{(i)}|\hat H|n_1^{(j)},n_2^{(j)},...,n_M^{(j)}\rangle,
	\label{e.38}
	\end{equation}
	where $i,j\in\{0,1,...,\mathrm n-1\}$ label the $\mathrm n$ Fock states.  Using the algorithm presented herein, we then define the elements $\mathrm a_{i,j}$ for the matrix representation of the Hamiltonian $\hat{\mathrm H}$ and calculate the coefficients $\mathrm c_{r_\mathcal Q}$, which we subsequently use as input in the VQE algorithm.  This algorithm requires that the operator is of the form of Eq.\,(\ref{e.11}).

	As a demonstration of this approach, we performed a set of transformations and VQE calculations for a relativistic system containing four massless interacting bosons in a box of dimension $2\times1\times1$ in units of $L$.  We restricted the number of modes to two; specifically we let $M_x=2$ and $M_y=M_z=1$.  As $N=4$ and $M=2$, the dimension given by Eq.\,(\ref{e.37}) of the Hilbert space $\mathcal V_\mathrm n$ for our chosen system is $\mathrm n=15$.  The Hamiltonian $\hat H$ operating on $\mathcal V_\mathrm n$ has $57$ nonzero matrix elements, which we obtained from Eq.\,(\ref{e.38}) for various coupling constants $\lambda$.  Next, we injected $\mathcal V_\mathrm n$ into the $16$-dimensional Hilbert space for a quantum register $\mathcal H^\mathrm Q$ with $\mathrm Q=4$ qubits.  For this injection, we chose $\Delta=100$ arb. units, which is in principle large enough to prevent the fictitious state introduced by the injection from: (1) becoming the ground state associated with the Hamiltonian $\hat{\mathrm H}$ on $\mathcal H^\mathrm Q$, and (2) appreciably affecting the physical states of our system.  In reality, choosing a large $\Delta$ has the drawback that the noise present in quantum computing calculations, which introduces weights into fictitious states, results to an artificial increase in the calculated ground-state energies.
	
	We applied the above transformation to produce the Pauli coordinates describing $\hat H$ for various values of $\lambda$, which we subsequently used as input to the VQE algorithm.  We then executed this algorithm using the Aer, Aqua, and Ignis application programming interfaces included in IBM Qiskit~\cite{qiskit}.  We chose a variational form of the quantum algorithm with a quantum circuit consisting of RY single-qubit gates and CNOT entangling gates repeated with the depth of four~\cite{Fischer19}.  We optimized the RY rotations using the StatevectorSimulator and the Powell optimizer in Qiskit with the convergence criteria that the relative error in the solution is less than $10^{-4}$.  Lastly, we calculated the final ground-state energies independently on the simulator and the IBM Q Santiago quantum computer.  For each circuit, we averaged the output over 8,192 identical runs to reduce statistical noise.
	
	\begin{figure}[t]
		\centering
		\includegraphics{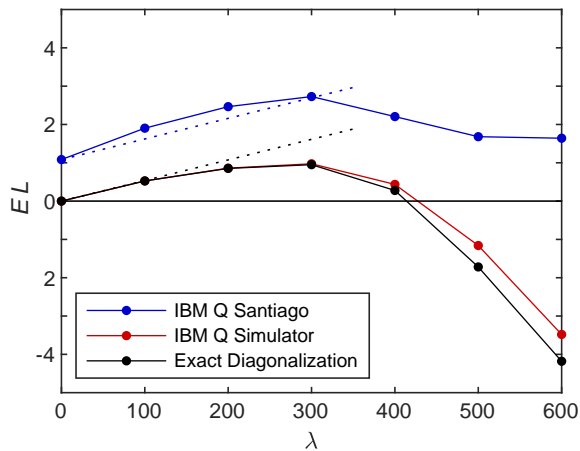}
		\caption{Scaled ground-state energy $EL$ in natural units of four massless relativistic bosons in a two-mode box of dimensions $2\times1\times1$ in units of length $L$ as a function of the dimensionless interaction coupling constant $\lambda$.  The energy has been obtained from exact diagonalization of the Hamiltonian, as well as from the VQE algorithm executed on an IBM Q Simulator and the IBM Q Santiago quantum computer.  The dotted lines show the approximate scaled energy $E_\mathrm{approx}L=E_0L+\beta\lambda L$ for weak interactions with the slope $\beta\approx5.374\times10^{-3}$ obtained from perturbation theory.}
		\label{f.4}
	\end{figure}
	Figure~\ref{f.4} shows the final ground-state energy for different interaction strengths calculated independently on the simulator and the IBM Q Santiago quantum computer.  Also shown is the ground-state energy obtained from exact diagonalization of $\hat{\mathrm H}$.  Though we observe appreciable absolute errors in the quantum computing calculations, the overall shape of the curve is similar to that of the exact diagonalization curve, suggesting that the quantum computing calculations at least produce a reasonable approximation of the ground state.
	
	Let us now explore the ground-state energy for weak interactions.  The ground-state energy $E$ can then be approximated by the first-order perturbation energy $E^{(1)}=\langle0|\hat V|0\rangle$, which we can express as
	\begin{equation}
	E^{(1)} = \frac{3\lambda}{4!}\frac{(\hbar c)^3}{4\Omega}\sum_{\vec\mu\vec\xi}\frac{1}{\hbar\omega_{\vec\mu}\,\hbar\omega_{\vec\xi}}\prod_\alpha\bigg(1+\frac{\delta_{\mu_\alpha,\xi_\alpha}}{2}\bigg).
	\end{equation}
	Note that this expression is independent of $N$, suggesting that the interacting ground-state energy depends only weakly on the maximum allowed number of particles in the box.  Irrespective of $N$, there are exactly two modes $(1,1,1)$ and $(2,1,1)$ in our considered system, which we label $1$ and $2$, respectively, for short.  The ground-state energy for our two-mode interacting system is then approximately
	\begin{equation}
	E^{(1)} = \frac{3\lambda}{4!}\frac{(\hbar c)^3}{64L}\bigg[\frac{27}{(\hbar\omega_1L)^2}+\frac{27}{(\hbar\omega_2L)^2}+\frac{16}{(\hbar\omega_1L)(\hbar\omega_2L)}\bigg],
	\end{equation}
	where the scaled mode energies from Eq.\,(\ref{e.32}) are $\hbar\omega_1L=3\pi\hbar c/2\approx4.712$ and $\hbar\omega_2L=\sqrt{3}\pi\hbar c\approx5.441$, in natural units.  The approximate scaled ground-state energy for weak interactions can thus be expressed $E^{(1)}L=\beta\lambda$, where the slope $\beta\approx5.374\times10^{-3}$.  As shown in Fig.~\ref{f.4}, $\beta\lambda$ provides an excellent approximation to the exact ground-state energy for weak interactions.  The corresponding estimated slope from our first two data points calculated on IBM Q Santiago is $\beta_\mathrm{qc}\approx8.192\times10^{-3}$.  Although by no means a perfect estimate, the comparison shows that the quantum computer we used already has sufficient accuracy to provide ballpark prediction.
	
	\section{Conclusions}
	\label{s.5}
	
	One of the challenges facing quantum computing applications is the conformation of classical input data to the input requirements of quantum algorithms such as the VQE and the HHL algorithms.  As the preprocessing algorithm present herein requires no more than $\mathcal O(\mathrm N^2\log_2\mathrm N)$ arithmetic operations, it could in conjunction with a quantum or hybrid quantum/classical algorithm offer an overall speedup over purely classical linear algebra algorithms requiring $\mathcal O(\mathrm N^3)$ arithmetic operations.  Further speedup is still possible for specific linear operators describing systems exhibiting some form of symmetry.  However, the real limitation we ran into for large $\mathrm N$ was not compute time, but rather available memory.  The number of matrix elements or coordinates needed to describe a general linear operator is $\mathrm N^2$.  We therefore expect that in the foreseeable future, quantum computing applications will continue to be most powerful for problems requiring large calculations with limited amount of input data.  Even so, our hope is that the preprocessing algorithm herein will nevertheless make a broader class of problems accessible to quantum computing.
%=============================================================================%
	\begin{acknowledgments}
		This work has been supported by the Office of Naval Research (ONR) through the U.S. Naval Research Laboratory (NRL) and by NRL through a Naval Innovative Science and Engineering (NISE) program.  We acknowledge quantum computing resources from IBM through a collaboration with the Air Force Research Laboratory (AFRL).
	\end{acknowledgments}
%=============================================================================%
%

%=============================================================================%
\end{document}